\documentstyle[12pt,fleqn]{article}

\title{Power densities for two-step gamma-ray transitions\\
 from isomeric states}
\author{Silviu Olariu and Agata Olariu\\
Institute of Physics and Nuclear Engineering,\\
Department of Fundamental Experimental Physics\\
76900 Magurele, P.O. Box MG-6, Bucharest, Romania\\
e-mail: olariu@ifin.nipne.ro}

\begin{document}
\date{}
\maketitle

\abstract
We have calculated the incident photon power density $P_2$ for which the 
two-step 
induced emission rate from an isomeric nucleus becomes equal to the natural 
isomeric decay rate. We have analyzed two-step transitions for isomeric 
nuclei with a half-life greater than 10 min, for which there is an 
intermediate state of known energy, spin and half-life, for which the 
intermediate state is connected by a known gamma-ray transition to the 
isomeric state and to at least another intermediate state, and for which 
the relative intensities of the transitions to lower states are known.
For the isomeric nucleus $^{166m}$Ho, which has a 1200 y isomeric 
state at 5.98 keV, we have found a value of $P_2=6.3 \times 10^7$ 
W cm$^{-2}$, the intermediate state being the 263.8 keV 
level. We have found power densities $P_2$ of the order of $10^{10}$ 
W cm$^{-2}$ for several other isomeric nuclei. \\

\endabstract
\newpage

The induced deexcitation of an isomeric nucleus is a two-step process in which 
the nucleus,  
initially in the isomeric state $|i\rangle$, first absorbs a photon of energy
$E_{ni}$ to reach a higher intermediate state $|n\rangle$, then
the nucleus makes a transition to a lower state $|l\rangle$ by the emission
of a gamma-ray photon of energy  $E_{nl}$ or by internal conversion.
In a previous work \cite{1} it has been required that the cascade originating 
on the state $|n\rangle$ should contain a gamma-ray transition of energy 
$E_\gamma >2 E_{ni}$, the latter relation representing 
a condition of upconversion of the energy of the incident photons,  
and it was found that in favorable cases the two-step induced emission rates 
become equal to the natural isomeric decay rates for incident power densities 
of the order of $10^{10}$ W cm$^{-2}$. The Weisskopf estimates where used
in \cite{1} in cases when tabulated half-lives or partial widths were not 
available.

In this report we have analyzed, regardless of the upconversion condition 
$E_\gamma >2 E_{ni}$, the two-step transitions for isomeric nuclei
with a half-life greater than 10 min, for which there is an intermediate state
$|n\rangle$ of known energy, spin and half-life, for which the intermediate 
state $|n\rangle$ is connected by a known gamma-ray transition to the isomeric 
state and to a lower state $|l\rangle$ , 
and for which the relative intensities
of the transitions from the intermediate state $|n\rangle$ 
are known. \cite{2} We have calculated the 
incident power density $P_2$ for which the two-step induced emission rate 
becomes equal to the natural isomeric decay rate.  We have found a value of 
$P_2=6.3 \times 10^7$ W cm$^{-2}$ for $^{166}$Ho, which has a 1200 y isomeric 
state at $E_i$=5.98 keV, and the intermediate state is the $E_n$=263.8 keV 
level, so that $E_{ni}=257.8$ keV. We have also found several isomers for
which the power density $P_2$ is of the order of $10^{10}$ W cm$^{-2}$. 

We assume that the nucleus is initially in an isomeric state $|i\rangle$ of
energy $E_i$, spin $J_i$ and half-life $t_i$.
By absorbing an incident photon  of energy
$E_{ni}$, the nucleus makes a transition to a higher
intermediate state $|n\rangle$ of energy $E_n$, spin $J_n$ and half-life $t_n$.
The state $|n\rangle$ then decays into a lower
state $|l\rangle$ of energy $E_l$ and spin $J_l$ 
by the emission of a gamma-ray photon having the  energy
$E_{nl}$ or by internal conversion, 
as shown in Fig. 1. In some cases the state $|l\rangle$ may be
situated above the isomeric state $|i\rangle$, and in these cases the
transition $|n\rangle\rightarrow |l\rangle$ is followed by further gamma-ray
transitions to lower states.

As shown in \cite{1}, the rate $w_{il}^{(2)}$ for the two-step transition
$|i\rangle\rightarrow |n\rangle\rightarrow |l\rangle$ is
\begin{equation}
w_{il}^{(2)}=\sigma_{ni}\hbar\Gamma_{eff} N(E_{ni}) .
\label{1}
\end{equation}
where
\begin{equation}
\sigma_{ni}=\frac{2J_n+1}{2J_i+1}\frac{\pi^2 c^2\hbar^2}{E_{ni}^2} 
\label{2}
\end{equation}
is the induced-emission cross section for the transition $|i\rangle\rightarrow
|n\rangle$. The quantity $\Gamma_{eff}$ is the effective width of the 
two-step transition,
\begin{equation}
\Gamma_{eff}=F_R \ln 2/t_n ,
\label{3}
\end{equation} 
where the dimensionless quantity $F_R$ has the expression
\begin{equation}
F_R=\frac{(1+\alpha_{nl})R_{ni}R_{nl}}
{\left[(1+\alpha_{ni})R_{ni}+(1+\alpha_{nl})R_{nl}
+\sum_{l^\prime} (1+\alpha_{nl^\prime})R_{nl^\prime}\right]^2} \:,
\label{4}
\end{equation}
$R_{ni}, R_{nl}, R_{nl^\prime}$ being the relative gamma-ray intensities and 
$\alpha_{ni}, \alpha_{nl},  \alpha_{nl^\prime}$ 
the internal conversion coefficients
for the transitions $|n\rangle\rightarrow |i\rangle,
|n\rangle\rightarrow |l\rangle, |n\rangle\rightarrow |l^\prime \rangle, 
l^\prime\not=i,l.$  
In Eq. (\ref{1}) $N(E_{ni})$ is the
spectral intensity for the incident photon flux, defined such that $N(E)dE$ 
should represent the number of photons incident per unit surface and time 
and having the energy between $E$ and $E+dE$.

The spectral intensity $N_2$ for which the two-step transition rate 
$w_{il}^{(2)}$
becomes equal to the natural decay rate $\ln 2/t_i$ of the isomeric nucleus is
\begin{equation}
N_2=\frac{\ln 2}{\sigma_{int}t_i} ,
\label{5}
\end{equation}
where the integrated cross section $\sigma_{int}$ is
\begin{equation}
\sigma_{int}=\sigma_{ni}\hbar\Gamma_{eff} .
\label{6}
\end{equation}
The incident power density $P_2$ for which the induced emission rate becomes
equal to the natural decay rate of the isomeric state can then be estimated as
\begin{equation}
P_2=N_2(E_{ni}) E_{ni}^2. 
\label{7}
\end{equation}

We have analyzed two-step transitions for isomeric 
nuclei with a half-life greater than 10 min, for which there is an 
intermediate state $|n\rangle$ of known energy $E_n$, spin $J_n$ and 
half-life $t_n$, for which the 
intermediate state is connected by a known gamma-ray transition to the 
isomeric state $|i\rangle$ and to a lower state 
$|l\rangle$, and for 
which the relative intensities $R_{ni}, R_{nl}, R_{nl^\prime}$ 
of the transitions to lower states are known. The internal conversion 
coefficients have been calculated by interpolation 
from refs. \cite{3} and \cite{4}. The cases for
which all the previously mentioned quantities were available  in ref. \cite{2}
are listed in Table I. 
If the input values were available for several two-step
transitions of a certain isomeric nucleus, we gave the two-step
transition having the lowest value of $P_2$. The two-step transitions for 
the nuclei $^{52}$Mn, $^{99}$Tc, $^{152}$Eu, $^{178}$Hf, $^{201}$Bi, 
$^{204}$Pb, which fulfil the upconversion condition, have been studied 
in ref. \cite{1} and are not included in Table I. If the half-life
$t_n$ of the intermediate state was given in \cite{2} as less than $(<)$ or 
greater than $(>)$ a certain value,
we have calculated $\hbar \Gamma_{eff}, \: \sigma_{int} $ 
and $P_2$ corresponding to the given limiting value of $t_n$. In the case of 
$^{166}$Ho we have neglected the 3.1 keV transition, of unknown intensity, 
from the 263.8 keV level. In the case of $^{97}$Tc the intensity of the 441.2
keV transition from the 656.9 keV level was given as an upper limit. In the
case of $^{121}$Sn we have neglected the weak 56.35 keV transition
from the 925.6 keV level. The half-life of the 0.0768 keV 
isomeric level of $^{235}$U and the intensity
of the 637.7 keV transition from the 637.79 keV intermediate state are
approximate values. In the case of $^{34}$Cl the intensity of the
2433.8 keV transition from the 2580.2 keV intermediate state was given as an
upper limit, as are the intensities of some other transitions from this
intermediate state. In the case of $^{123}$Sn we have neglected the weak 284.7
keV transition from the 1130.5 keV intermediate state. 

For the isomeric nucleus $^{166m}$Ho, which has a 1200 y isomeric 
state at $E_i$=5.98 keV, we have found a value of $P_2=6.3 \times 10^7$ 
W cm$^{-2}$, the intermediate state being the $E_n$=263.8 keV 
level. The relatively low value of $P_2$ is due to the long half-life of this
isomer. 
The largest effective width in Table I is for $^{34m}$Cl, for which
$\hbar\Gamma_{eff} = 2.5\times 10^{-3}$ eV. The relatively large effective
width is due to the short half-life of the intermediate state, $t_n$=3 fs.
The energy of the pumping transition is however large, $E_{ni}$=2433.8 keV, 
and the multipolarity is E2/M1.
The large value of $P_2$ for $^{34}$Cl is due to the relatively small half-life
of the isomeric state. There are several cases in Table I for which 
the power density $P_2$ is of the order of $10^{10}$ W cm$^{-2}$. For $^{97}$Tc
and $^{95}$Tc the low values of $P_2$ can be attributed to the 
short half-lives of the intermediate state, 
whereas for $^{113}$Cd and $^{121}$Sn  the values of $P_2$ 
are a result of the long isomeric half-lives.
The two-step transitions for which all the input values 
were available represent a small fraction of the total
number of possible two-step transitions in isomeric nuclei.

Since the number of isomeric nuclei in a sample is limited by the total
activity of that sample, a longer isomeric half-life means a larger number of
isomeric nuclei in the sample. The two-step induced transitions can be induced
by  incident photons, as assumed in this work,  
or directly by incident electrons.  The cross sections for the two-step
electron
excitation of isomeric states are about two orders of magnitudes lower than the
cross sections for two-step photon excitation but 
the two-step photon excitation rates are also proportional to the 
efficiency with which the energy of an incident electron beam can be converted 
into bremsstrahlung.

\newpage

Fig. 1. Two-step deexcitation of an isomeric nucleus. 
The nucleus, initially in the isomeric state $|i\rangle$, makes a transition
to a higher nuclear state $|n\rangle$ by the absorption of an incident photon 
of energy $E_{ni}$, then it emits a photon of energy $E_{nl}$ 
to reach the lower nuclear state $|l\rangle$.
\vspace*{1in}

Table I. Power density $P_2$ for which the two-step induced emission rate 
becomes equal to the natural decay rate of the isomeric state of energy $E_i$
and half-life $t_i$. The intermediate state of the problem is the level of 
energy  $E_n$ and half-life $t_n$. The multipolarities of the transitions of 
energy $E_{ni}, E_{nl}$ are given in the multipole column.

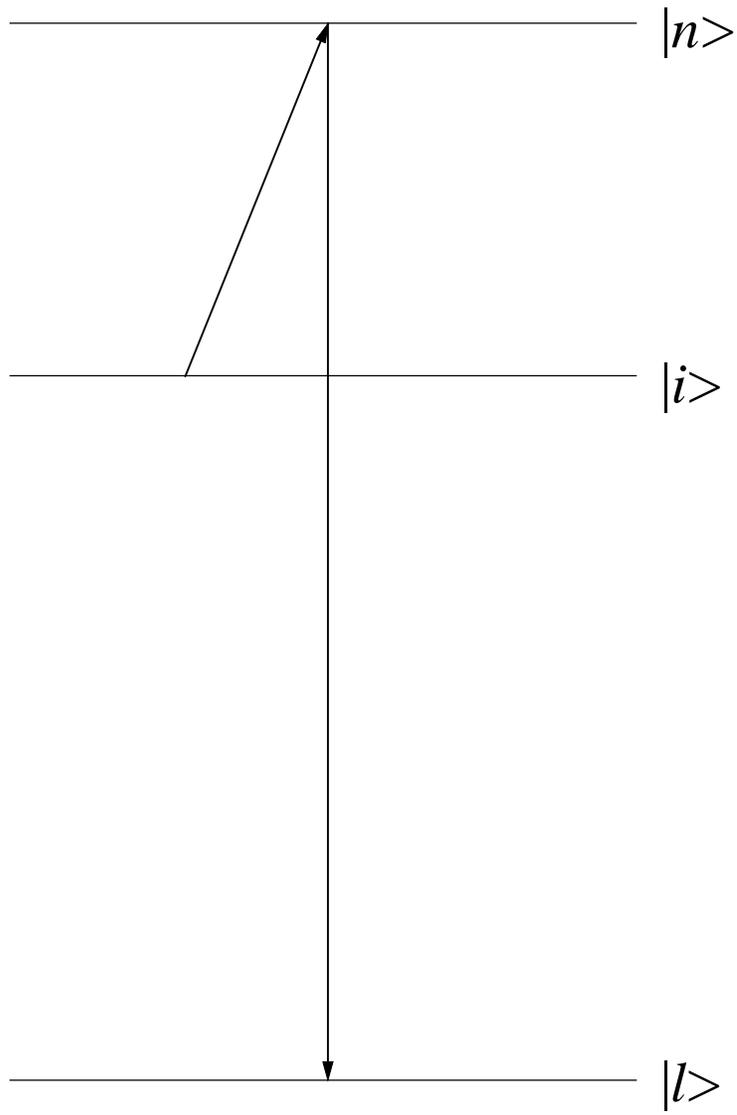
\end{document}